\documentclass[11pt,a4paper]{article}

\usepackage{jheppub}
\usepackage[T1]{fontenc}
\usepackage[utf8]{inputenc}
\usepackage{bm}
\usepackage{mathtools}
\usepackage{mathrsfs}
\usepackage{xcolor}

\allowdisplaybreaks

\newcommand{\ii}{\mathrm{i}}
\newcommand{\dd}{\mathrm{d}}
\newcommand{\e}{\mathrm{e}}
\newcommand{\order}{\mathcal{O}}
\newcommand{\field}{\bm{\phi}}
\newcommand{\mode}{\mathbf{U}}
\newcommand{\identity}{\mathbf{1}}

\newcommand{\im}{\operatorname{Im}}
\newcommand{\re}{\operatorname{Re}}

\title{Cosmological Collider Signals at Strong Mixing}

\author[a,b]{Xiangwei Wang,}
\author[a,b]{Yi Wang,}
\author[a,b]{Yunke Zhao}

\emailAdd{xwanglh@connect.ust.hk}
\emailAdd{phyw@ust.hk}
\emailAdd{yzhaogt@connect.ust.hk}

\affiliation[a]{Department of Physics, The Hong Kong University of Science and
Technology, Clear Water Bay, Kowloon, Hong Kong S.A.R., P.R. China}
\affiliation[b]{Jockey Club Institute for Advanced Study, The Hong Kong
University of Science and Technology, Hong Kong S.A.R., P.R. China}

\abstract{We study cosmological collider signatures in a two-field
inflationary system with constant-turn derivative mixing between the
canonically normalized curvature fluctuation and a massive isocurvature
field. Building on recently derived exact hypergeometric solutions that treat
the quadratic mixing nonperturbatively, we construct the mixed propagators constituting the cosmological correlators. A Mellin-Barnes approach helps isolate the pair of nonanalytic
late-time branches carried by a heavy field. For a cubic isocurvature
self-interaction, these branches contribute towards a squeezed bispectrum with a
power-law envelope and logarithmic oscillations. The oscillation frequency
encodes the heavy mass, while the amplitude and phase retain the full mixing
dependence rather than a perturbative expansion. We perform the squeezed limit integral
analytically and
compare it with numerical results across
representative masses and mixing strengths. At fixed mixing we derive the
large-mass JWKB expansion of 
the squeezed correlator and its phase. We also give integral representations for
the isocurvature Wightman function. The research
idea was suggested by ARC, the calculation was performed by GPT, and the
results were checked by the authors.}


\begin{document}
\maketitle

\section{Introduction}
\label{sec:introduction}

Inflation turns primordial correlation functions into probes of degrees of
freedom at energies that are otherwise difficult to access.  Cosmological
collider physics pursues this possibility through the nonanalytic momentum
dependence generated by additional fields with masses of order the Hubble
scale.  Early studies of massive fields coupled to the curvature perturbation
through a turning trajectory established the mixing between the curvature perturbation and the isocurvature perturbation, the
interpolation between local and nonlocal bispectrum shapes, and the
mass-dependent squeezed scaling $(k_L/k_S)^{3/2-\nu}$~\cite{Chen:2009we,Chen:2009zp,Baumann:2011nk,Chen:2012ge,Pi:2012gf,Achucarro:2012sm,Noumi:2012vr,Arkani-Hamed:2015bza}.

For a heavy field in the principal series, the two late-time modes become oscillatory.  The evolution of their oscillatory quantum phase,
equivalently its frequency in $\ln(k_L/k_S)$, encodes the mass, whereas angular
dependence encodes the spin.  A relative phase or phase offset between
contributions is a separate observable containing quantum-interference
information.  Massive fields can also act as quantum primordial standard
clocks whose oscillatory non-Gaussianity records the background
evolution~\cite{Chen:2015lza}.  Effective-field-theory analyses
have classified the couplings of massive spinning fields to the Goldstone mode
and the graviton and analyzed the scalar and tensor correlators that they
induce beyond the conformally invariant squeezed-limit setup~\cite{Lee:2016vti}.
 These results isolate a universal
nonanalytic structure, but the amplitude and phase with which it is transferred
to curvature perturbations remain sensitive to the inflationary dynamics.

In perturbative treatments of turning-trajectory systems,
curvature-isocurvature transfer is often expanded in the turn rate.
Schwinger-Keldysh diagrammatics and mixed propagators provide an efficient
organization of this expansion, including in the presence of derivative
mixing~\cite{Chen:2017ryl}.
When the turn is rapid, however, repeated quadratic mixing cannot be truncated
reliably.  Earlier work clarified the validity of effective single-field
descriptions in strongly coupled two-field systems~\cite{Cremonini:2010ua},
studied the onset of the large-mixing effective theory with analytic and
numerical methods~\cite{An:2017hlx}, unified approximate large-mass and
large-mixing limits~\cite{Tong:2017iat}, and developed a partial effective
theory that retains nonlocal clock signals~\cite{Iyer:2017qzw}.  Huenupi
et al.~\cite{Huenupi:2026abj} recently went beyond these approximations by
deriving exact analytic solutions of the coupled linear curvature-isocurvature
system for arbitrary dimensionless mixing $\lambda$ and isocurvature mass $\mu$ in
quasi-de Sitter space.  They also derived the curvature power spectrum in
closed form.  Higher-point functions were explicitly left there as a
future application of the exact linear basis.

Analytic studies of inflationary correlators have meanwhile used Mellin-Barnes representations to expose their pole and nonanalytic
structure~\cite{Sleight:2019hfp,Sleight:2019mgd}, obtained closed-form expressions for broad
classes of massive exchange correlators~\cite{Qin:2023ejc}, and organized
nested integrals with multiple massive exchanges through partial
Mellin-Barnes representations~\cite{Xianyu:2023ytd}.  Here the
Mellin-Barnes representation is applied directly to the exact strongly mixed
linear basis in order to isolate the late-time branches of the mixed external
leg.

In this work we make that extension for a cubic isocurvature self-interaction
$g_{\sigma^3}\sigma^3$ with the quadratic derivative mixing treated exactly while
working to first order in the cubic coupling.  The two exact Bunch-Davies
solutions determine the Wightman functions and subsequently the external leg propagators
$K_{Aq}=G^>_{Aq}(\tau,0)$, where $A\in\{q,\sigma\}$, representing one of the two fields in our model.

Using the method of ref.~\cite{Huenupi:2026abj}, we obtain
the mixed external leg, which effectively resums the derivative mixing.  A Mellin-Barnes
representation then separates analytic late-time terms from the massive
nonanalytic branches
$z^{3/2\pm\ii\rho}$, where $\rho=-i\nu=\sqrt{\frac{\mu^2}{H^2}-\frac94}$.  For the squeezed configuration
$(k_1,k_2,k_3)=(k,k,ck)$, $c\ll1$, the latter give
\begin{equation}
 B_{q,\mathrm{nonan}}^{(\sigma^3)}(k,k,ck)
 =-2g_{\sigma^3}\frac{H^2}{k^6}
 \frac{|C_{\lambda,\rho}J_{\lambda,\rho}|}{c^{3/2}}
 \sin\!\left(\rho\ln c+\theta_{\lambda,\rho}\right).
 \label{eq:intro-squeezed-result}
\end{equation}
The logarithmic frequency is fixed by the heavy mass, while the exact soft
coefficient $C_{\lambda,\rho}$ and hard integral $J_{\lambda,\rho}$ retain the
nonperturbative mixing dependence of the amplitude and phase.

The exact solutions also make the fixed-mixing large-mass limit
tractable.  The soft coefficient admits direct gamma-function asymptotics,
whereas a JWKB analysis of the hard integral must be matched to its endpoint
Bessel layer.  At fixed $\lambda$ this gives a controlled expansion through
relative order $1/\rho$.  The leading algebraic contribution comes from the
positive-cycle endpoint rather than from a finite complex saddle.

The paper is organized as follows.  Section~\ref{sec:system} introduces the
strongly mixed two-field system, the propagator
dictionary, and the exact power spectrum.  Section~\ref{sec:exact-external}
constructs the exact mixed external leg and extracts its late-time
nonanalyticity.  In section~\ref{sec:bispectrum} we compute the tree-level
bispectrum from the cubic isocurvature interaction and derive its squeezed
oscillations. Section~\ref{sec:large-mass} demonstrates the large-mass expansion, and section~\ref{sec:discussion} discusses the
results and future directions. Technical details of the Bunch-Davies branch
sum, the Mellin-Barnes representation, the analytic hard-integral series, the bulk-to-bulk propagators and the
fixed-mixing JWKB expansion are collected in the appendices.

\paragraph{Note added.}
While this manuscript was being completed, we became aware of two concurrent works, Refs.~\cite{HuenupiEtAl:2026bispectrum,LucasEtAl:2026strongmixing}, addressing potentially related topics. We thank the authors of Refs.~\cite{HuenupiEtAl:2026bispectrum,LucasEtAl:2026strongmixing} for coordinating their arXiv submissions with us.

\section{The strongly mixed two-field system}
\label{sec:system}

We work at leading order in the de Sitter approximation, set the reduced
Planck mass to one, and take $H$, $\lambda$, and $\mu$ to be constant.  The
slow-roll parameter $\epsilon$ is retained in the conversion between the
canonically normalized curvature fluctuation
\begin{equation}
 q \equiv \zeta_c = \sqrt{2\epsilon}\,\zeta
 \label{eq:q-definition}
\end{equation}
and the observable curvature perturbation $\zeta$.  The isocurvature fluctuation is
denoted by $\sigma$.  For a constant turn rate $\Omega_{\rm turn}$, we define
\begin{equation}
 \lambda=-\frac{2\Omega_{\rm turn}}{H}.
 \label{eq:lambda-definition}
\end{equation}
The exact solution used below treats $\lambda$ nonperturbatively and follows
the strong-mixing construction of ref.~\cite{Huenupi:2026abj}.

\subsection{Quadratic action and equations of motion}

The quadratic action is
\begin{equation}
 S_2=\int \dd^4x\,a^3\left[
 \frac12\dot q^2-\lambda H\dot q\,\sigma
 +\frac12\lambda^2H^2\sigma^2
 -\frac{(\bm\nabla q)^2}{2a^2}
 +\frac12\dot\sigma^2
 -\frac{(\bm\nabla\sigma)^2}{2a^2}
 -\frac12\mu^2\sigma^2\right].
 \label{eq:quadratic-action}
\end{equation}
It is useful to collect the fields into
\begin{equation}
 \field=\begin{pmatrix}q\\ \sigma\end{pmatrix},\qquad
 \Gamma=\begin{pmatrix}0&-\lambda H\\0&0\end{pmatrix},\qquad
 D_t=\partial_t+\Gamma,
 \label{eq:connection}
\end{equation}
and to introduce
\begin{equation}
 \widetilde M^2=\begin{pmatrix}0&0\\0&\mu^2\end{pmatrix}.
 \label{eq:mass-matrix}
\end{equation}
Then
\begin{equation}
 S_2=\frac12\int\dd^4x\,a^3\left[
 (D_t\field)^{\mathsf T}(D_t\field)
 -\frac{1}{a^2}(\bm\nabla\field)^{\mathsf T}
     (\bm\nabla\field)
 -\field^{\mathsf T}\widetilde M^2\field\right].
 \label{eq:compact-action}
\end{equation}
For a Fourier mode, the exact matrix equation following from
eq.~\eqref{eq:compact-action} is
\begin{equation}
 \frac{1}{a^3}\frac{\dd}{\dd t}
 \left(a^3D_t\field_{\bm k}\right)
 -\Gamma^\dagger D_t\field_{\bm k}
 +\left(\frac{k^2}{a^2}+\widetilde M^2\right)\field_{\bm k}=0.
 \label{eq:matrix-eom-covariant}
\end{equation}
Before specializing to constant parameters, its expanded form contains the
connection derivative,
\begin{equation}
 \ddot\field_{\bm k}+3H\dot\field_{\bm k}
 +(\Gamma-\Gamma^\dagger)\dot\field_{\bm k}
 +\left(\dot\Gamma+3H\Gamma-\Gamma^\dagger\Gamma
 +\frac{k^2}{a^2}+\widetilde M^2\right)\field_{\bm k}=0.
 \label{eq:matrix-eom-expanded}
\end{equation}
In the constant-$H$, constant-$\lambda$ limit, $\dot\Gamma=0$, and the two
component equations become
\begin{align}
 \ddot q_{\bm k}+3H\dot q_{\bm k}+\frac{k^2}{a^2}q_{\bm k}
 -\lambda H\dot\sigma_{\bm k}-3\lambda H^2\sigma_{\bm k}&=0,
 \label{eq:q-eom}\\
 \ddot\sigma_{\bm k}+3H\dot\sigma_{\bm k}
 +\left(\frac{k^2}{a^2}+\mu^2-\lambda^2H^2\right)\sigma_{\bm k}
 +\lambda H\dot q_{\bm k}&=0.
 \label{eq:sigma-eom}
\end{align}

At zero momentum the first equation can be written as
\begin{equation}
 \frac{\dd}{\dd t}\left[a^3(\dot q-\lambda H\sigma)\right]=0.
 \label{eq:superhorizon-constraint}
\end{equation}
After the decaying integration constant is discarded,
$\dot q=\lambda H\sigma$.  Substitution into
eq.~\eqref{eq:sigma-eom} removes the apparent $-\lambda^2H^2$ mass shift and
gives the familiar late-time indices
\begin{equation}
 \sigma\sim z^{3/2+\nu},\ z^{3/2-\nu},\qquad
 z\equiv\frac{k}{aH}=-k\tau.
 \label{eq:late-indices}
\end{equation}

\subsection{Bunch-Davies modes and canonical normalization}

There are two independent positive-frequency solutions, labeled by
$b=1,2$.  We arrange them as the columns of the mode matrix
\begin{equation}
 \mode_k(z)=
 \begin{pmatrix}
  q_1(k,z)&q_2(k,z)\\
  \sigma_1(k,z)&\sigma_2(k,z)
 \end{pmatrix}.
 \label{eq:mode-matrix}
\end{equation}
The field operator is expanded as
\begin{equation}
 \field_{\bm k}(t)=\mode_k(t)\,\bm a_{\bm k}
 +\mode_k^*(t)\,\bm a^\dagger_{-\bm k}.
 \label{eq:operator-expansion}
\end{equation}
Since the canonical momentum is
$\bm\Pi_{\bm k}=a^3D_t\field_{\bm k}$, the equal-time commutator is
equivalent to the field-space Wronskian condition
\begin{equation}
 a^3\left[
 \mode_k(D_t\mode_k)^\dagger
 -\mode_k^*(D_t\mode_k)^{\mathsf T}\right]=\ii\identity.
 \label{eq:wronskian}
\end{equation}

For the exact solution, it is convenient to decompose each branch in a free
mode basis,
\begin{align}
 q_b(k,z)&=\zeta_b^+(z)u_0(k,z)+\zeta_b^-(z)u_0^*(k,z),
 \label{eq:q-bogoliubov}\\
 \sigma_b(k,z)&=\sigma_b^+(z)u_\mu(k,z)
 +\sigma_b^-(z)u_\mu^*(k,z),
 \label{eq:sigma-bogoliubov}
\end{align}
where $\zeta_b^\pm$ are Bogoliubov coefficients for the canonical field $q$.  The properly normalized basis is
\begin{align}
 u_0(k,z)&=\ii\frac{H}{\sqrt{2k^3}}(1-\ii z)\e^{\ii z},
 \label{eq:u0}\\
 u_\mu(k,z)&=-\frac{H}{\sqrt{2k^3}}\sqrt{\frac{\pi}{2}}
 \e^{\ii\theta_\nu}z^{3/2}H_\nu^{(1)}(z),
 \qquad \theta_\nu=\frac{\pi}{2}\left(\nu-\frac32\right).
 \label{eq:umu}
\end{align}
Here $\theta_\nu$ chosen to impose the Bunch-Davies phase. 

\subsection{Wightman functions and the external legs}
\label{sec:kernel-dictionary}

The two Wightman functions are
\begin{align}
 G^>_{AB}(k;\tau_1,\tau_2)
 &=\sum_{b=1}^2 U_{Ab}(k,\tau_1)U^*_{Bb}(k,\tau_2),
 \label{eq:wightman-greater}\\
 G^<_{AB}(k;\tau_1,\tau_2)
 &=G^>_{BA}(k;\tau_2,\tau_1),
 \label{eq:wightman-less}
\end{align}
where $A,B\in\{q,\sigma\}$.  The external leg ending on a
late-time curvature mode is
\begin{equation}
 K_{Aq}(k;\tau)
 \equiv G^>_{Aq}(k;\tau,0)
 =\sum_{b=1}^2U_{Ab}(k,\tau)q_b^*(k,0).
 \label{eq:external-leg}
\end{equation}
In particular,
\begin{equation}
 P_q(k)=K_{qq}(k;0)
 =\sum_{b=1}^2|q_b(k,0)|^2.
 \label{eq:Pq}
\end{equation}

Similarly we may construct the nonperturbative bulk-to-bulk propagators as 
\begin{align}
    G_{\sigma\sigma}^{>}(k;\tau_1,\tau_2)=\sum_{b=1}^2\sigma_b(k,\tau_1)\sigma^*_b(k,\tau_2)\,.
\end{align}
see appendix~\ref{app:kernel} for details. 

\subsection{Power spectrum}

Direct mode quantization gives
\begin{equation}
 P_\zeta(k)=\frac{P_q(k)}{2\epsilon},\qquad
 \Delta_\zeta(k)=\frac{k^3}{2\pi^2}P_\zeta(k)
 =\frac{k^3}{4\pi^2\epsilon}\sum_{b=1}^2|q_b(k,0)|^2.
 \label{eq:power-spectrum}
\end{equation}
Writing the single-field result as
\begin{equation}
 \Delta_0=\frac{H^2}{8\pi^2\epsilon},
 \label{eq:Delta0}
\end{equation}
the exact linear solution of ref.~\cite{Huenupi:2026abj} gives the
nonperturbative power-spectrum ratio
\begin{equation}
 \frac{\Delta_\zeta}{\Delta_0}=
 \left|
 \frac{
 \Gamma\left(\frac34-\frac\nu2\right)
 \Gamma\left(\frac34+\frac\nu2\right)}{
 \Gamma\left(\frac34-\frac\nu2+\frac{\ii\lambda}{2}\right)
 \Gamma\left(\frac34+\frac\nu2+\frac{\ii\lambda}{2}\right)}
 \right|^2.
 \label{eq:exact-power-ratio}
\end{equation}

\section{The mixed external leg from exact linear modes}
\label{sec:exact-external}

The first-order Bogoliubov system permits the isocurvature mode to be reconstructed
from the positive-frequency coefficient of $q$.  With the conventions in
eq.~\eqref{eq:u0}, the corrected identities are
\begin{align}
 \sigma_b(k,z)&=-\frac{\sqrt{2}H}{\lambda k^{3/2}}
 z^2\e^{\ii z}\frac{\dd\zeta_b^+(z)}{\dd z},
 \label{eq:sigma-reconstruction}\\
 q_b^*(k,0)&=-\ii\frac{H}{\sqrt{2k^3}}
 \left[\zeta_b^+(0)-\zeta_b^-(0)\right]^*.
 \label{eq:q-boundary}
\end{align}
Together these relations reconstruct the exact
$K_{\sigma q}=\sum_b\sigma_bq_b^*$.

\subsection{Exact special-function basis}

Following the exact linear construction of ref.~\cite{Huenupi:2026abj}, define
the hypergeometric differential operator
\begin{equation}
 \widehat{\mathscr D}_{\nu,s}
 ={}_2F_1\left(\frac12-\nu,\frac12+\nu;
 1+\ii s\lambda;\frac{\ii}{2}\frac{\dd}{\dd z}\right),
 \qquad s=\pm1,
 \label{eq:Dhat}
\end{equation}
understood through its formal power series, and set
\begin{equation}
 a_s=\frac{\ii s\lambda}{2},\qquad
 f_s(z)=\widehat{\mathscr D}_{\nu,s}
 U(a_s,1,-2\ii z),
 \label{eq:fs-definition}
\end{equation}
where $U(a,b,z)$ is the Tricomi function. 
Summing the two Bunch-Davies branches gives the exact external leg
\begin{equation}
 K_{\sigma q}(k;z)=\frac{H^2}{k^3}T_{\sigma q}(z),
 \label{eq:transfer-definition}
\end{equation}
where
\begin{equation}
 T_{\sigma q}(z)=\frac{\ii}{2\lambda}z^2\e^{\ii z}
 \sum_{s=\pm1}\e^{s\pi\lambda/2}\mathcal U_s^*f_s'(z),
 \label{eq:exact-transfer}
\end{equation}
and the late-time coefficient is
\begin{equation}
 \mathcal U_s=\frac{1}{\sqrt\pi}
 \frac{
 \Gamma\left(\frac12+a_s\right)
 \Gamma\left(\frac34-\frac\nu2\right)
 \Gamma\left(\frac34+\frac\nu2\right)}{
 \Gamma\left(\frac34-\frac\nu2+a_s\right)
 \Gamma\left(\frac34+\frac\nu2+a_s\right)}.
 \label{eq:Us-definition}
\end{equation}

\subsection{Late-time nonanalyticity}
\label{sec:late-transfer}

The Mellin-Barnes representation derived in appendix~\ref{app:MB} separates
the analytic terms from the two nonanalytic late-time branches.  For generic
$\nu$,
\begin{equation}
 T_{\sigma q}(z)=C^+_{\lambda,\nu}z^{3/2+\nu}
 +C^-_{\lambda,\nu}z^{3/2-\nu}+\order(z^2).
 \label{eq:T-late-general}
\end{equation}
The coefficient of the first branch is
\begin{equation}
 C^+_{\lambda,\nu}=-\frac{1}{\lambda}
 \sum_{s=\pm1}\left[
 \e^{\ii\pi\phi_s/2}2^{\nu-1/2}
 \frac{\Gamma(-2\nu)\Gamma(1+2a_s)}{
 \Gamma(a_s)\Gamma\left(2a_s-\nu+\frac12\right)}
 \mathcal U_s^*\right],
 \qquad
 \phi_s=\frac12-\nu-\ii s\lambda,
 \label{eq:Cplus}
\end{equation}
and $C^-_{\lambda,\nu}$ follows from the same expression under
$\nu\to-\nu$. 

For the heavy case $\rho=-i\nu\in\mathbb{R}$, we may define
\begin{equation}
 C_{\lambda,\rho}\equiv C^+_{\lambda,\ii\rho}=(C^-_{\lambda,\ii\rho})^*
 ,
\end{equation}
such that
\begin{equation}
 T_{\sigma q}(z)=C_{\lambda,\rho}z^{3/2+\ii\rho}
 +C_{\lambda,\rho}^*z^{3/2-\ii\rho}+\order(z^2).
 \label{eq:T-late-heavy}
\end{equation}
These two complex
powers carry the nonanalytic mass dependence of the cosmological collider
signal.

\section{Cubic non-Gaussianity}
\label{sec:bispectrum}

We consider the isocurvature self-interaction
\begin{equation}
 \mathcal L_{\rm int}\supset-\frac{a^4}{3!}g_{\sigma^3}\sigma^3.
 \label{eq:cubic-interaction}
\end{equation}
The bispectrum convention is
\begin{equation}
 \langle q_{\bm k_1}q_{\bm k_2}q_{\bm k_3}\rangle
 =(2\pi)^3\delta^{(3)}(\bm k_1+\bm k_2+\bm k_3)
 B_q(k_1,k_2,k_3).
 \label{eq:B-convention}
\end{equation}
At tree level the exact in-in result is written solely in terms of the external legs of eq.~\eqref{eq:external-leg},
\begin{equation}
 B_q^{(\sigma^3)}(k_1,k_2,k_3)
 =-2g_{\sigma^3}\im\int_{-\infty}^0
 \dd\tau\,a^4(\tau)
 \prod_{i=1}^3K_{\sigma q}(k_i;\tau).
 \label{eq:exact-bispectrum}
\end{equation}

Choose a reference momentum $k_*$ and define
\begin{equation}
 r_i=\frac{k_i}{k_*},\qquad x=-k_*\tau.
 \label{eq:dimensionless-variables}
\end{equation}
Using $a=-1/(H\tau)$ and eq.~\eqref{eq:transfer-definition}, the exact
bispectrum becomes
\begin{equation}
 B_q^{(\sigma^3)}(k_1,k_2,k_3)
 =-2g_{\sigma^3}\frac{H^2}{k_*^6}
 \frac{1}{r_1^3r_2^3r_3^3}
 \im\,\mathcal I_{\lambda,\rho}(r_1,r_2,r_3),
 \label{eq:B-dimensionless}
\end{equation}
where
\begin{equation}
 \mathcal I_{\lambda,\rho}(r_1,r_2,r_3)
 =\int_0^\infty\dd x\,x^{-4}
 \prod_{i=1}^3T_{\sigma q}(r_ix).
 \label{eq:I-bispectrum}
\end{equation}

\subsection{The squeezed limit}

Let $k$ be the hard momentum and $ck$ the soft momentum, with $c\ll1$.
Taking $k_*=k$ gives
\begin{equation}
 B_q^{(\sigma^3)}(k,k,ck)
 =-2g_{\sigma^3}\frac{H^2}{k^6c^3}
 \im\int_0^\infty\dd x\,x^{-4}
 T_{\sigma q}(x)^2T_{\sigma q}(cx).
 \label{eq:B-squeezed-integral}
\end{equation}
The nonanalytic part of the soft transfer function is
\begin{equation}
 T_{\sigma q}(cx)=C_{\lambda,\rho}(cx)^{3/2+\ii\rho}
 +C_{\lambda,\rho}^*(cx)^{3/2-\ii\rho}
 +\text{analytic terms}.
 \label{eq:soft-transfer}
\end{equation}
We define the two hard integrals
\begin{align}
 J^+_{\lambda,\rho}&=\int_0^\infty\dd x\,
 x^{-5/2+\ii\rho}T_{\sigma q}(x)^2,
 \label{eq:Jplus}\\
 J^-_{\lambda,\rho}&=\int_0^\infty\dd x\,
 x^{-5/2-\ii\rho}T_{\sigma q}(x)^2,
 \label{eq:Jminus}
\end{align}
and the combination
\begin{equation}
 J_{\lambda,\rho}
 \equiv J^+_{\lambda,\rho}-(J^-_{\lambda,\rho})^*
 =2\ii\int_0^\infty\dd x\,x^{-5/2+\ii\rho}
 \im\left[T_{\sigma q}(x)^2\right].
 \label{eq:J-combination}
\end{equation}
The plus sign in the last equality follows directly from
$T^2-(T^*)^2=2\ii\im(T^2)$.  

Writing
\begin{equation}
 \theta_{\lambda,\rho}=\arg\left(
 C_{\lambda,\rho}J_{\lambda,\rho}\right),
 \label{eq:theta}
\end{equation}
the nonanalytic squeezed signal takes the compact form
\begin{equation}
 \boxed{
 B_{q,\mathrm{nonan}}^{(\sigma^3)}(k,k,ck)
 =-2g_{\sigma^3}\frac{H^2}{k^6}
 \frac{|C_{\lambda,\rho}J_{\lambda,\rho}|}{c^{3/2}}
 \sin\left(\rho\ln c+\theta_{\lambda,\rho}\right).}
 \label{eq:B-squeezed-final}
\end{equation}
Equation~\eqref{eq:B-squeezed-final} displays the characteristic logarithmic
oscillation, while the exact transfer function retains the full dependence on
the mixing.  Analytic soft terms have been omitted from the displayed signal.
For the observable curvature perturbation,
\begin{equation}
 B_\zeta=\frac{B_q}{(2\epsilon)^{3/2}}.
 \label{eq:Bzeta-conversion}
\end{equation}

\subsection{Analytic hard integral and numerical comparison}
\label{sec:J-series-comparison}

For real $\lambda>0$, $\rho>0$, and $\eta=\pm1$, identify
$J^{\eta=+1}=J^+$ and $J^{\eta=-1}=J^-$, and define
\begin{equation}
 p_\eta=\frac52+\eta\ii\rho,\quad
 \alpha_\eta=\frac12-\eta\ii\rho,\quad d_{s\eta}=\alpha_\eta+2a_s,
 \quad
 \omega_s=\frac{\e^{s\pi\lambda/2}\mathcal U_s^*}{\Gamma(a_s)}.
 \label{eq:J-series-parameters}
\end{equation}
The two continuous integrations in the hard coefficient can be performed
exactly.  The resulting one-leg moment is
\begin{equation}
 I_{s,n}^{(\eta)}
 =B(n+a_s+1,2)
 {}_3F_2\!\left(
 \begin{matrix}
  \alpha_\eta,\ d_{s\eta},\ n+a_s+1\\
  1+2a_s,\ n+a_s+3
 \end{matrix};1\right),
 \label{eq:J-series-moment}
\end{equation}
and the exact one-fold series is
\begin{equation}
 \displaystyle
 J^\eta_{\lambda,\rho}
 =\frac{\Gamma(p_\eta)}{(-2\ii)^{p_\eta}\lambda^2}
 \sum_{n=0}^{\infty}\frac{(p_\eta)_n}{n!}
 \left[\sum_{s=\pm1}\omega_s I_{s,n}^{(\eta)}\right]^2.
 \label{eq:J-analytic-series}
\end{equation}
  The series is absolutely convergent and is combined into the physical
$J_{\lambda,\rho}$ according to eq.~\eqref{eq:J-combination}.  Its derivation is given in
appendix~\ref{app:J-series}.  The $\lambda\to0$ limit must be taken only after
the two branches have been combined.

\section{Fixed-mixing large-mass limit}
\label{sec:large-mass}

We now take $\rho\to\infty$ with
$\lambda>0$ fixed.  Stirling expansion of the exact soft coefficient gives
\begin{equation}
 \begin{aligned}
 C_{\lambda,\rho}
 ={}&\frac{\e^{-\pi\rho}\sinh(\pi\lambda/2)}{\sqrt{\rho}}
 \exp\!\left\{\ii\left[
 \frac{\pi}{2}-\rho\bigl(\ln (2\rho)-1\bigr)\right]\right\}
 \\[-1mm]
 &\times\left[1+\frac{\ii(6\lambda^2+1)}{12\rho}
 +\order_\lambda(\rho^{-2})\right].
 \end{aligned}
 \label{eq:C-large-rho}
\end{equation}
The JWKB waves of the hard kernel must be matched to the conical endpoint at
$t=0$.  The resulting endpoint expansion is
\begin{equation}
 \begin{aligned}
 J_{\lambda,\rho}
 ={}&\frac{\sqrt{\pi}\,\lambda^2}{16\rho^2}
 \exp\!\left\{\ii\left[
 \rho\left(\ln \frac{\rho}{2}-1\right)+\frac{3\pi}{4}
 \right]\right\}
 \\[-1mm]
 &\times\left[1+\frac{\ii(12\lambda^2+1)}{24\rho}
 +\order_\lambda(\rho^{-2})\right].
 \end{aligned}
 \label{eq:J-large-rho}
\end{equation}
Combining the two expansions consistently through relative order $1/\rho$
gives
\begin{equation}
 \begin{aligned}
 C_{\lambda,\rho}J_{\lambda,\rho}
 ={}&\frac{\sqrt{\pi}\,\lambda^2\sinh(\pi\lambda/2)}
 {16\rho^{5/2}}\e^{-\pi\rho}
 \exp\!\left\{\ii\left(\frac{5\pi}{4}-2\rho\ln 2\right)\right\}
 \\[-1mm]
 &\times\left[1+\frac{\ii(\lambda^2+\frac18)}{\rho}
 +\order_\lambda(\rho^{-2})\right],
 \end{aligned}
 \label{eq:CJ-large-rho}
\end{equation}
and hence
\begin{equation}
 \boxed{
 \theta_{\lambda,\rho}
 =\frac{5\pi}{4}-2\rho\ln 2
 +\frac{\lambda^2+\frac18}{\rho}
 +\order_\lambda(\rho^{-2})\quad(\bmod\ 2\pi).}
 \label{eq:theta-large-rho}
\end{equation}
The expansion is controlled at fixed $\lambda$; a conservative practical
condition for the displayed correction is $\lambda^2/\rho\ll1$.  Apparently the Boltzmann suppression may disappear when $\lambda^2/\rho $ is of order $\mathcal{O}(1)$, but this is beyond the validity of our approximation made here, and is interesting to be carefully examined in the future. The details are given in appendix~\ref{app:large-rho}.

\begin{figure}[tbp]
 \centering
 \includegraphics[width=0.82\textwidth]{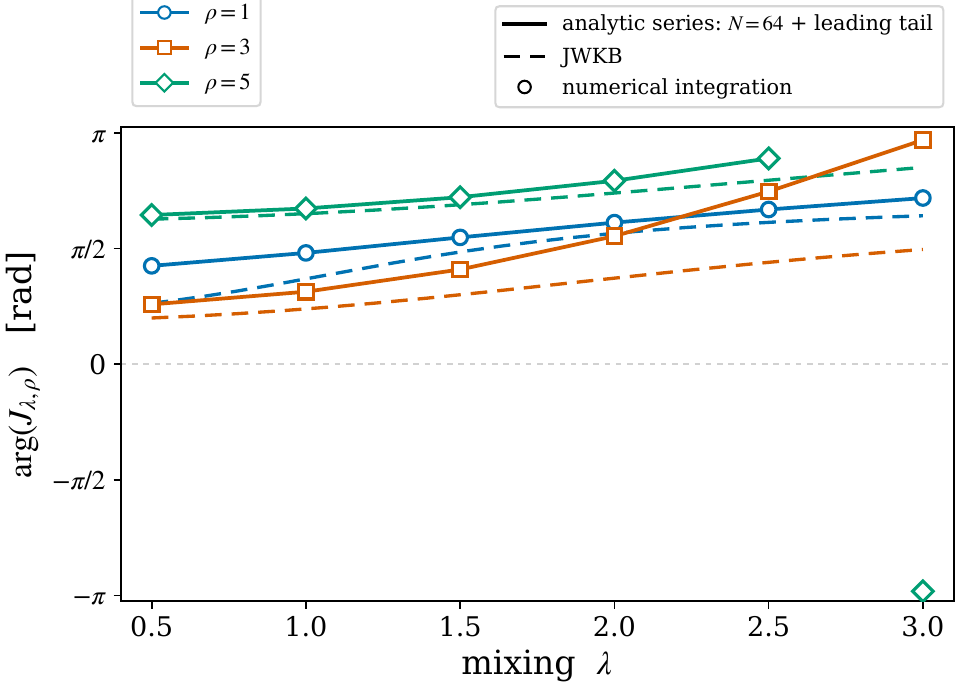}
 \caption{Principal phase $\operatorname{Arg}J_{\lambda,\rho}$ for
 $\rho=1,3,5$.  Solid lines denote the analytic $N=64$ series, including the
 leading endpoint tail, while points denote the full double-integral
 evaluation using an independent finite-cutoff numerical integration.  Dashed
 lines show the JWKB result~\eqref{eq:J-large-rho}.  Phases take
 values in $(-\pi,\pi]$; solid and dashed lines are broken whenever adjacent
 values cross the principal branch cut.}
 \label{fig:J-arg-comparison}
\end{figure}

\begin{figure}[tbp]
 \centering
 \includegraphics[width=0.82\textwidth]{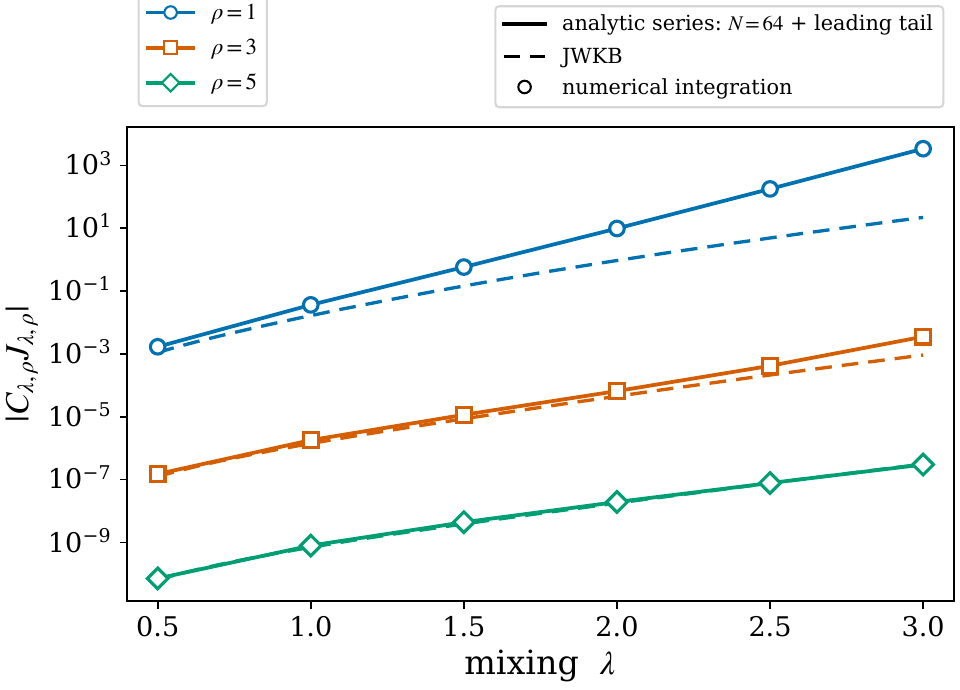}
 \caption{Magnitude $|C_{\lambda,\rho}J_{\lambda,\rho}|$ controlling the
 nonanalytic squeezed amplitude in eq.~\eqref{eq:B-squeezed-final}.  Solid
 lines use the analytic $N=64$ hard-integral series with its leading endpoint
 tail and the exact coefficient $C_{\lambda,\rho}$, while points denote the full double-integral evaluation using an
independent finite-cutoff numerical integration.
 Dashed lines show the combined JWKB approximation~\eqref{eq:CJ-large-rho}. The grid is $\rho=1,3,5$, and the vertical axis is
 logarithmic.}
 \label{fig:CJ-abs-comparison}
\end{figure}

\begin{figure}[tbp]
 \centering
 \includegraphics[width=0.82\textwidth]{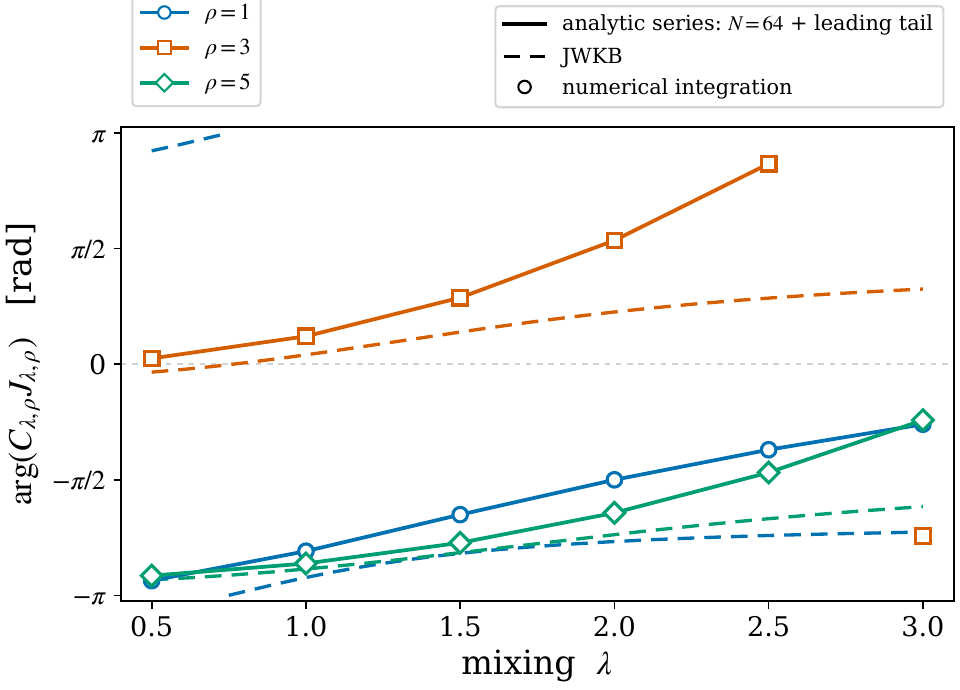}
 \caption{Principal squeezed-signal phase
 $\theta_{\lambda,\rho}=\operatorname{Arg}
 (C_{\lambda,\rho}J_{\lambda,\rho})$.  Solid lines use the analytic $N=64$
 series with its leading endpoint tail, while points denote the full double-integral evaluation using an
independent finite-cutoff numerical integration.  Dashed
 lines show the JWKB phase following from
 eq.~\eqref{eq:CJ-large-rho}.  The grid is $\rho=1,3,5$.  Values lie in
 $(-\pi,\pi]$; solid and dashed lines are broken whenever adjacent values cross the principal
branch cut.}
 \label{fig:CJ-arg-comparison}
\end{figure}

Figures~\ref{fig:J-abs-comparison}-\ref{fig:CJ-arg-comparison} compare the
series truncated at $N=64$, an independent finite-cutoff numerical integration, and the
JWKB approximation on the common grid $\rho=1,3,5$.  The dashed
curves are shown outside as well as inside their formal asymptotic regime to
display directly where the large-$\rho$ approximation becomes reliable.

\begin{figure}[tbp]
 \centering
 \includegraphics[width=0.82\textwidth]{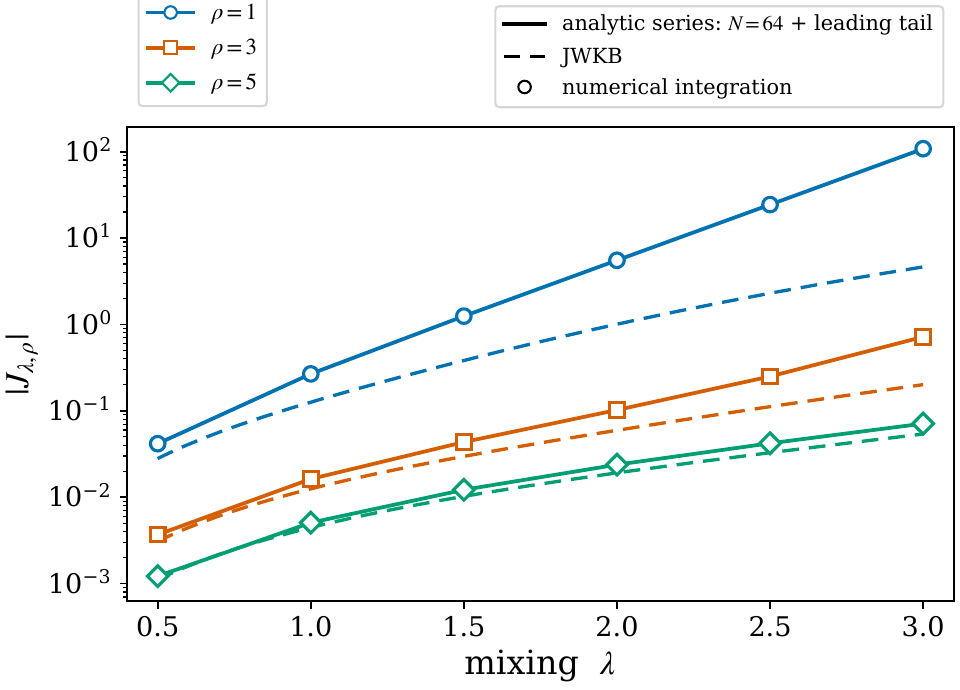}
 \caption{Hard-integral magnitude for $\rho=1,3,5$.  Solid lines show the
 analytic series~\eqref{eq:J-analytic-series} truncated at $N=64$ and supplied
 with the leading endpoint tail described in appendix~\ref{app:J-series}.
 Points show the full double-integral evaluation using an independent
 finite-cutoff numerical integration, and dashed lines show the JWKB
 approximation~\eqref{eq:J-large-rho}.  The vertical axis is logarithmic.}
 \label{fig:J-abs-comparison}
\end{figure}

\clearpage

\section{Discussion}
\label{sec:discussion}

The exact linear modes and closed-form curvature power spectrum used here were
derived by Huenupi et al.~\cite{Huenupi:2026abj}.  Building on that linear
solution, we have constructed the mixed external leg needed for higher-point
in-in correlators and applied it to the cubic isocurvature interaction. 

For the cubic isocurvature interaction, the exact mixed external leg first reduces
the calculation to a single time integral, and the analytic construction in
appendix~\ref{app:J-series} then removes the remaining continuous integrations
in favor of an absolutely convergent series.  The late-time nonanalyticity
fixes the frequency and scaling of the squeezed signal, while the hard
coefficient determines the mixing-dependent amplitude and phase.  The JWKB method is applied to obtain more explicit result in the large $\rho $ limit. In this approximation, we see possibilities of lifting the Boltzmann suppression on the signal, when $\lambda^2/\rho \sim1$, which deserves further studies. 

\acknowledgments

We thank Xi Tong, Zhong-Zhi Xianyu, and Yong Sheng Yap for helpful discussions.  We thank the authors of refs.~\cite{HuenupiEtAl:2026bispectrum,LucasEtAl:2026strongmixing} for coordinating the submissions. We used ARC, described in
ref.~\cite{ma2026arc}
(\href{https://chinaxiv.org/abs/202606.00234}{paper}) and available from its
\href{https://github.com/tririver/arc/}{GitHub repository}, 
for literature-domain construction, research-idea suggestion, and part of the calculations. This work is
supported in part by the RGC Research Fellow Grant RFS2425-6S02 from the
Research Grants Council of Hong Kong.

\clearpage
\appendix

\section{Bunch-Davies branch sum}
\label{app:BD-sum}

This appendix follows ref.~\cite{Huenupi:2026abj} and records the branch sum underlying
eq.~\eqref{eq:exact-transfer}.  At a large initial value $z_0$, the two
independent Bunch-Davies solutions have the Tricomi-branch coefficients
\begin{align}
 A_+^{(1)}&=\frac12\e^{\pi\lambda/4}
 \e^{\frac{\ii\lambda}{2}\ln(2z_0)}+\order(z_0^{-1}),
 &
 A_-^{(1)}&=\frac12\e^{-\pi\lambda/4}
 \e^{-\frac{\ii\lambda}{2}\ln(2z_0)}+\order(z_0^{-1}),
 \label{eq:A-branch-one}\\
 A_+^{(2)}&=-\frac{\ii}{2}\e^{\pi\lambda/4}
 \e^{\frac{\ii\lambda}{2}\ln(2z_0)}+\order(z_0^{-1}),
 &
 A_-^{(2)}&=\frac{\ii}{2}\e^{-\pi\lambda/4}
 \e^{-\frac{\ii\lambda}{2}\ln(2z_0)}+\order(z_0^{-1}).
 \label{eq:A-branch-two}
\end{align}
The Kummer-$M$ branches vanish in the Bunch-Davies limit.  Thus the part of
$\zeta_b^+$ relevant for the exact external leg is
\begin{equation}
 \zeta_b^+(z)=A_+^{(b)}f_+(z)+A_-^{(b)}f_-(z)+\order(z_0^{-1}).
 \label{eq:zeta-branch-expansion}
\end{equation}
The branch sums obey
\begin{align}
 \sum_{b=1}^2|A_+^{(b)}|^2&=\frac12\e^{\pi\lambda/2},
 &
 \sum_{b=1}^2|A_-^{(b)}|^2&=\frac12\e^{-\pi\lambda/2},
 \label{eq:A-diagonal-sums}\\
 \sum_{b=1}^2A_+^{(b)}A_-^{(b)*}&=0.
 \label{eq:A-cross-sum}
\end{align}
Combining these identities with eqs.~\eqref{eq:sigma-reconstruction} and
\eqref{eq:q-boundary} gives
\begin{equation}
 \sum_{b=1}^2\sigma_b(k,z)q_b^*(k,0)
 =\frac{\ii H^2}{2\lambda k^3}z^2\e^{\ii z}
 \sum_{s=\pm1}\e^{s\pi\lambda/2}\mathcal U_s^*f_s'(z),
 \label{eq:external-leg-branch-sum}
\end{equation}
which is precisely eq.~\eqref{eq:exact-transfer}.

\section{The mixed propagator and its Mellin-Barnes representations}
\label{app:MB}

The Tricomi integral
\begin{equation}
 U(a,b,w)=\frac{1}{\Gamma(a)}\int_0^\infty\dd t\,
 \e^{-wt}t^{a-1}(1+t)^{b-a-1}
 \label{eq:Tricomi-integral}
\end{equation}
is initially valid for $\re a>0$ and $|\arg w|<\pi/2$.  In the present
application $a_s=\ii s\lambda/2$ lies on the boundary of this domain.
Equations below therefore mean the analytic continuation selected by the
Bunch-Davies $\ii\epsilon$ prescription.

Acting with the hypergeometric differential operator gives
\begin{equation}
 f_s(z)=\frac{1}{\Gamma(a_s)}\int_0^\infty\dd t\,
 \e^{2\ii zt}t^{a_s-1}(1+t)^{-a_s}
 {}_2F_1\left(\frac12-\nu,\frac12+\nu;1+2a_s;-t\right).
 \label{eq:fs-integral}
\end{equation}
For $\nu=\ii\rho$, differentiation and the standard associated-Legendre
normalization give
\begin{align}
 f_s'(z)
 &=\frac{2\ii}{\Gamma(a_s)}\int_0^\infty\dd t\,
 \e^{2\ii zt}t^{a_s}(1+t)^{-a_s}
 {}_2F_1\left(\frac12-\ii\rho,\frac12+\ii\rho;
 1+2a_s;-t\right)
 \notag\\
 &=\frac{2\ii\Gamma(1+2a_s)}{\Gamma(a_s)}
 \int_0^\infty\dd t\,\e^{2\ii zt}
 P_{\ii\rho-1/2}^{-2a_s}(1+2t)
 \notag\\
 &=\frac{\ii\Gamma(1+2a_s)}{\Gamma(a_s)}
 \int_1^\infty\dd y\,\e^{\ii z(y-1)}
 P_{\ii\rho-1/2}^{-2a_s}(y).
 \label{eq:fs-Legendre}
\end{align}
The factor $\Gamma(1+2a_s)$ in the last two lines follows from
\begin{equation}
 t^{a_s}(1+t)^{-a_s}
 {}_2F_1\left(\frac12-\ii\rho,\frac12+\ii\rho;
 1+2a_s;-t\right)
 =\Gamma(1+2a_s)P_{\ii\rho-1/2}^{-2a_s}(1+2t).
 \label{eq:Legendre-normalization}
\end{equation}

For the late-time expansion, use
\begin{align}
 (1+t)^{-a_s}
 &=\frac{1}{2\pi\ii\Gamma(a_s)}\int_{C_u}\dd u\,
 \Gamma(a_s+u)\Gamma(-u)t^u,
 \label{eq:MB-binomial}\\
 {}_2F_1\left(\frac12-\nu,\frac12+\nu;1+2a_s;-t\right)
 &=\frac{1}{2\pi\ii}
 \frac{\Gamma(1+2a_s)}{
 \Gamma\left(\frac12-\nu\right)\Gamma\left(\frac12+\nu\right)}
 \notag\\[-2mm]
 &\quad\times\int_{C_v}\dd v\,
 \frac{
 \Gamma\left(\frac12-\nu+v\right)
 \Gamma\left(\frac12+\nu+v\right)\Gamma(-v)}{
 \Gamma(1+2a_s+v)}t^v.
 \label{eq:MB-hypergeometric}
\end{align}
The contours separate the left-going and right-going pole families.  After the
$t$ integral,
\begin{equation}
 f_s'(z)=\frac{1}{2\ii\pi^2}
 \frac{\Gamma(1+2a_s)}{
 \Gamma\left(\frac12-\nu\right)
 \Gamma\left(\frac12+\nu\right)\Gamma(a_s)^2}
 \mathscr I_s(z),
 \label{eq:fs-MB}
\end{equation}
where
\begin{equation}
 \mathscr I_s(z)=\int_{C_u}\dd u\int_{C_v}\dd v\,
 \Phi_L(\nu,a_s;u,v)\Phi_R(u,v)
 (-2\ii z)^{-(u+v+a_s+1)},
 \label{eq:Is-MB}
\end{equation}
with
\begin{align}
 \Phi_L(\nu,a_s;u,v)
 &=\frac{
 \Gamma(a_s+u)
 \Gamma\left(\frac12-\nu+v\right)
 \Gamma\left(\frac12+\nu+v\right)
 \Gamma(u+v+a_s+1)}{
 \Gamma(1+2a_s+v)},
 \label{eq:PhiL}\\
 \Phi_R(u,v)&=\Gamma(-u)\Gamma(-v).
 \label{eq:PhiR}
\end{align}

Closing the contour to the left, we obtain
\begin{align}
 \mathscr I_s(z)
 &=(2\pi\ii)^2\sum_{m,n\geq0}
 \frac{(-1)^{m+n}}{m!\,n!}
 \bigg[
 \mathcal R^{(+)}_{m,n}(-2\ii z)^{m+n+\nu-1/2}
 \notag\\[-1mm]
 &\hspace{16mm}
 +\mathcal R^{(-)}_{m,n}(-2\ii z)^{m+n-\nu-1/2}+\mathcal{S}^{(1)}_{m,n}(-2iz)^m+\mathcal{S}^{(2)}_{m,n} (-2iz)^n
 \bigg],
\end{align}
where
\begin{align}
 \mathcal R^{(+)}_{m,n}
 &=\frac{
 \Gamma(-n-2\nu)
 \Gamma\left(-m-n-\nu+\frac12\right)}{
 \Gamma\left(-n+2a_s-\nu+\frac12\right)}
 \Gamma\left(n+\nu+\frac12\right)\Gamma(m+a_s),
 \label{eq:Rplus}\\
 \mathcal R^{(-)}_{m,n}
 &=\frac{
 \Gamma(-n+2\nu)
 \Gamma\left(-m-n+\nu+\frac12\right)}{
 \Gamma\left(-n+2a_s+\nu+\frac12\right)}
 \Gamma\left(n-\nu+\frac12\right)\Gamma(m+a_s),\label{eq:Rminus}\\
\mathcal{S}_{m,n}^{(1)}&=\sum_{r=\pm}\frac{\Gamma\left(-m+n-\frac{1}{2}-r\nu\right)\Gamma\left(2r\nu-n\right)}{\Gamma\left(-n+2a_s+\frac{1}{2}+r\nu\right)}\nonumber\\
 &~~~~~~~~~~~~~~~~~~~~~\times\Gamma\left(n+\frac{1}{2}-r\nu\right)\Gamma\left(m-n+a_s+\frac{1}{2}+r\nu\right),\\
 \mathcal{S}_{m,n}^{(2)}&=\frac{\Gamma\left(m-n-\nu-\frac{1}{2}\right)\Gamma\left(m-n+\nu-\frac{1}{2}\right)}{\Gamma\left(m-n+2a_s\right)}\Gamma(m+a_s)\Gamma(-m+n+1)\,.
\end{align}
The two families of nonanalytic residues yield
\begin{align}
 \mathscr I_s^{\rm nonan}(z)
 &=(2\pi\ii)^2\sum_{m,n\geq0}
 \frac{(-1)^{m+n}}{m!\,n!}
 \bigg[
 \mathcal R^{(+)}_{m,n}(-2\ii z)^{m+n+\nu-1/2}
 \notag\\[-1mm]
 &\hspace{16mm}
 +\mathcal R^{(-)}_{m,n}(-2\ii z)^{m+n-\nu-1/2}
 \bigg].
 \label{eq:I-nonanalytic-residues}
\end{align}
The remaining pole families generate integer powers of $z$.  Keeping
$m=n=0$ in eq.~\eqref{eq:I-nonanalytic-residues} and multiplying by the
prefactors in eqs.~\eqref{eq:exact-transfer} and \eqref{eq:fs-MB} produces
eqs.~\eqref{eq:T-late-general} and \eqref{eq:Cplus}.

\section{The Schwinger-Keldysh bulk-to-bulk kernel}
\label{app:kernel}

The Schwinger-Keldysh bulk-to-bulk kernel is defined as
\begin{align}
    G_{\sigma\sigma}^{>}(k;\tau_1,\tau_2) = \sum_{b=1}^2 \sigma_b(k,\tau_1)\,\sigma^*_b(k,\tau_2), \label{eq:A1}
\end{align}
where the individual mode functions are given by
\begin{equation}
    \sigma_b(k,z)=-\frac{\sqrt{2}H}{\lambda k^{\frac32}}z^2e^{iz}\frac{\mathrm{d}\zeta_b^+}{\mathrm{d}z}. \label{eq:A2}
\end{equation}
Here $z=-k\tau$ and $k$ is the internal momentum. Using the explicit relations among the amplitudes, the sum over the derivatives of \(\zeta_b^+\) reduces to
\begin{equation}
    \sum_{b=1}^2
    \frac{\mathrm{d}\zeta^+_b(z_1)}{\mathrm{d}z_1}
    \frac{\mathrm{d}\zeta^{+*}_b(z_2)}{\mathrm{d}z_2}
    =
    \frac{e^{\pi\lambda/2}}{2}f'_1(z_1)f'^*_1(z_2)
    +
    \frac{e^{-\pi\lambda/2}}{2}f'_2(z_1)f'^*_2(z_2).
\label{eq:A4}
\end{equation}
Substituting this into the definition yields the greater and lesser Green functions
\begin{align}
    G_{\sigma\sigma}^>(k;z_1,z_2) &=
    \frac{2H^2}{\lambda^2k^3}z_1^2z_2^2e^{i(z_1-z_2)}
    \left(\frac{e^{\pi\lambda/2}}{2}f'_1(z_1)f'^*_1(z_2)+\frac{e^{-\pi\lambda/2}}{2}f'_2(z_1)f'^*_2(z_2)\right), \label{eq:A5}\\
    G_{\sigma\sigma}^<(k;z_1,z_2) &=
    \frac{2H^2}{\lambda^2k^3}z_1^2z_2^2e^{-i(z_1-z_2)}
    \left(\frac{e^{\pi\lambda/2}}{2}f'^*_1(z_1)f'_1(z_2)+\frac{e^{-\pi\lambda/2}}{2}f'^*_2(z_1)f'_2(z_2)\right). \label{eq:A6}
\end{align}
With the late-time asymptotic behavior of the mode functions,
the non-analytic momentum dependence of the kernel takes the form
\begin{equation}
    G_{\sigma\sigma}(k;\tau_1,\tau_2)
    \sim
    C_1 k^{2i\rho}
    +
    C_2 k^{-2i\rho}
    +
    C_3 .
\label{eq:A8}
\end{equation}
For our purposes it is more convenient to rewrite the sum in terms of conformal time. After expanding in the small-\(k\) limit one obtains
\begin{equation}
\begin{split}
\sum_{b=1}^2
\frac{\mathrm{d}\zeta^+_b(z_1)}{\mathrm{d}z_1}
\frac{\mathrm{d}\zeta^{+*}_b(z_2)}{\mathrm{d}z_2}
=
&\frac{\lambda^2}{k}(\tau_1\tau_2)^{-1/2}
\Big[
Q(4k^2\tau_1\tau_2)^{i\rho}
+
Q^*(4k^2\tau_1\tau_2)^{-i\rho}
\\
&+
P_1
\left(\frac{\tau_1}{\tau_2}\right)^{i\rho}
+
P_2
\left(\frac{\tau_1}{\tau_2}\right)^{-i\rho}
\Big]+\mathcal{O}\left(k^{-1/2}\right),
\label{eq:A9}
\end{split}
\end{equation}
with the coefficient
\begin{align}
    Q &= \frac12\,
    \frac{\Gamma(-2i\rho)^2}
         {\Gamma\bigl(\frac{1}{2}+i\lambda-i\rho\bigr)\Gamma\bigl(\frac{1}{2}-i\lambda-i\rho\bigr)}\,,\label{eq:A10}\\
         P_1&=\frac{\pi\,e^{\pi\rho}}{8\rho\cosh\left(\frac{\pi\lambda}{2}\right)\sinh(2\pi\rho)}
      \left(
        \frac{e^{\pi\lambda/2}}{\bigl|\Gamma\bigl(\tfrac12 + i\lambda - i\rho\bigr)\bigr|^2}
        +
        \frac{e^{-\pi\lambda/2}}{\bigl|\Gamma\bigl(\tfrac12 + i\lambda + i\rho\bigr)\bigr|^2}
      \right),\\
      P_2&=\frac{\pi\,e^{-\pi\rho}}{8\rho\cosh\left(\frac{\pi\lambda}{2}\right)\sinh(2\pi\rho)}
      \left(
        \frac{e^{-\pi\lambda/2}}{\bigl|\Gamma\bigl(\tfrac12 + i\lambda - i\rho\bigr)\bigr|^2}
        +
        \frac{e^{\pi\lambda/2}}{\bigl|\Gamma\bigl(\tfrac12 + i\lambda + i\rho\bigr)\bigr|^2}
      \right),
\end{align}
Finally, keeping only the leading non-analytic contribution,
the greater Green function can be written as
\begin{equation}
\begin{aligned}
    G_{\sigma\sigma}^>(k;\tau_1,\tau_2)
    = 2H^2(\tau_1\tau_2)^{3/2}\Bigl[ & (4\tau_1\tau_2)^{i\rho}Q\,e^{2i\rho\ln k} + (4\tau_1\tau_2)^{-i\rho}Q^*e^{-2i\rho\ln k} \\
    & + P_1\left(\frac{\tau_1}{\tau_2}\right)^{i\rho} + P_2\left(\frac{\tau_1}{\tau_2}\right)^{-i\rho}\Bigr] + \mathcal{O}(k^{1/2}).
\end{aligned}
\label{eq:A11}
\end{equation}
The spectral function is given by
\begin{equation}
    \begin{split}
        A(k;\tau_1,\tau_2)&=i\left(G_{\sigma\sigma}^>(k;\tau_1,\tau_2)-G_{\sigma\sigma}^<(k;\tau_1,\tau_2)\right)\\
        &=-\frac{H^2}{\rho}\left(\tau_1\tau_2\right)^{3/2}\sin\left(\rho\ln\left(\frac{\tau_1}{\tau_2}\right)\right).
    \end{split}
\end{equation}

\section{Analytic evaluation of the hard integral}
\label{app:J-series}

For real $\lambda>0$ and $\rho>0$, this appendix performs both continuous integrations in
$J^\eta_{\lambda,\rho}$ and derives the series quoted in
eq.~\eqref{eq:J-analytic-series}.  Throughout this appendix
$s,\eta\in\{-1,+1\}$ and the parameters are those in
eq.~\eqref{eq:J-series-parameters}.  The Bunch-Davies branch is fixed
explicitly by
\begin{equation}
 \ln (-2\ii)=\ln  2-\frac{\ii\pi}{2},\qquad
 (-2\ii)^{-p}=\exp\!\left[-p\left(\ln 2-\frac{\ii\pi}{2}\right)\right].
 \label{eq:J-log-branch}
\end{equation}

\subsection{Compactification and factorization}

Differentiating eq.~\eqref{eq:fs-integral}, define
\begin{align}
 h_s(t)&=t^{a_s}(1+t)^{-a_s}
 {}_2F_1\left(\frac12-\ii\rho,\frac12+\ii\rho;
 1+2a_s;-t\right),
 \label{eq:h-numerical}\\
 \mathcal H_{\lambda,\rho}(t)
 &=-\frac1\lambda\sum_{s=\pm1}\omega_s h_s(t).
 \label{eq:H-numerical}
\end{align}
Substitution into eq.~\eqref{eq:exact-transfer} gives
\begin{equation}
 T_{\sigma q}(x)=x^2\e^{\ii x}\int_0^\infty\dd t\,
 \e^{2\ii xt}\mathcal H_{\lambda,\rho}(t).
 \label{eq:T-H-numerical}
\end{equation}
Performing the $x$ integral on the regulated upper-half-plane contour first
gives
\begin{equation}
 J^\eta_{\lambda,\rho}
 =\frac{\Gamma(p_\eta)}{(-2\ii)^{p_\eta}}
 \int_0^\infty\dd t_1\int_0^\infty\dd t_2\,
 \frac{\mathcal H_{\lambda,\rho}(t_1)
       \mathcal H_{\lambda,\rho}(t_2)}{
 (1+t_1+t_2)^{p_\eta}}.
 \label{eq:J-double-integral}
\end{equation}
The two regulated branches are defined separately at this stage.  Their
discrete series may be combined only after eq.~\eqref{eq:J-double-integral}
has been evaluated with the prescription \eqref{eq:J-log-branch}.

Compactify each half-line according to
\begin{equation}
 t_j=\frac{x_j}{1-x_j},\qquad
 \dd t_j=\frac{\dd x_j}{(1-x_j)^2},\qquad
 1+t_1+t_2=\frac{1-x_1x_2}{(1-x_1)(1-x_2)}.
 \label{eq:J-compactification}
\end{equation}
Using the symmetry of the upper Gauss parameters, Pfaff's transformation on
the physical negative-real branch yields
\begin{equation}
 h_s\!\left(\frac{x}{1-x}\right)
 =x^{a_s}(1-x)^{\alpha_\eta}
 F_{s\eta}(x),\qquad
 F_{s\eta}(x)={}_2F_1(\alpha_\eta,d_{s\eta};1+2a_s;x).
 \label{eq:J-Pfaff-leg}
\end{equation}
The decisive simplification is
\begin{equation}
 p_\eta-2+\alpha_\eta=1.
 \label{eq:J-weight-cancellation}
\end{equation}
Consequently, eq.~\eqref{eq:J-double-integral} becomes
\begin{align}
 J^\eta_{\lambda,\rho}
 &=\frac{\Gamma(p_\eta)}{(-2\ii)^{p_\eta}\lambda^2}
 \sum_{s,r=\pm1}\omega_s\omega_r
 \int_0^1\!\dd x_1\int_0^1\!\dd x_2\,
 \frac{x_1^{a_s}x_2^{a_r}(1-x_1)(1-x_2)}
 {(1-x_1x_2)^{p_\eta}}
 \notag\\[-1mm]
 &\hspace{35mm}\times F_{s\eta}(x_1)F_{r\eta}(x_2).
 \label{eq:J-compact-double}
\end{align}

Expand the remaining kernel,
\begin{equation}
 (1-x_1x_2)^{-p_\eta}
 =\sum_{n=0}^\infty\frac{(p_\eta)_n}{n!}(x_1x_2)^n.
 \label{eq:J-kernel-series}
\end{equation}
One may first insert a factor $0<\xi<1$ as
$(1-\xi x_1x_2)^{-p_\eta}$, integrate term by term, and then take
$\xi\to1^-$ using the absolute convergence established below.  Each term
factorizes into the one-leg moment
\begin{equation}
 I_{s,n}^{(\eta)}
 =\int_0^1\dd x\,x^{n+a_s}(1-x)F_{s\eta}(x).
 \label{eq:J-one-leg-integral}
\end{equation}
The Euler beta-hypergeometric integral evaluates
eq.~\eqref{eq:J-one-leg-integral} as eq.~\eqref{eq:J-series-moment}. The standard terminating and balanced
${}_3F_2(1)$ summation conditions are not satisfied at generic
$(\lambda,\rho)$, so no finite gamma-only reduction is assumed.

\subsection{Absolute convergence and endpoint tail}

Put $\delta_\eta=2\eta\ii\rho$.  The connection formula at $x=1$ gives
\begin{align}
 F_{s\eta}(x)
 &=L_{s\eta}+M_{s\eta}(1-x)^{\delta_\eta}
 +\order(1-x)+\order\!\left((1-x)^{1+\delta_\eta}\right),
 \label{eq:J-endpoint-connection}\\
 L_{s\eta}
 &=\frac{\Gamma(1+2a_s)\Gamma(\delta_\eta)}
 {\Gamma(1+2a_s-\alpha_\eta)\Gamma(1+2a_s-d_{s\eta})},\\
 M_{s\eta}
 &=\frac{\Gamma(1+2a_s)\Gamma(-\delta_\eta)}
 {\Gamma(\alpha_\eta)\Gamma(d_{s\eta})}.
 \label{eq:J-endpoint-coefficients}
\end{align}
Beta integration then implies
\begin{equation}
 I_{s,n}^{(\eta)}=n^{-2}\left[
 L_{s\eta}+\Gamma(2+\delta_\eta)M_{s\eta}n^{-\delta_\eta}
 \right]+\order(n^{-3}).
 \label{eq:J-moment-asymptotic}
\end{equation}
Since $(p_\eta)_n/n!=\order(n^{3/2})$ in magnitude, the complete
summand in eq.~\eqref{eq:J-analytic-series} is $\order(n^{-5/2})$.
The series is therefore absolutely convergent and its bare remainder after
$n=0,\ldots,N-1$ is $\order(N^{-3/2})$.

For the acceleration used in the figures, define
\begin{equation}
 A_\eta=\sum_s\omega_sL_{s\eta},\qquad
 B_\eta=\Gamma(2+\delta_\eta)\sum_s\omega_sM_{s\eta}.
 \label{eq:J-tail-AB}
\end{equation}
The leading omitted tail is
\begin{align}
 \Delta_NJ^\eta
 &\sim\frac{(-2\ii)^{-p_\eta}}{\lambda^2}\bigg[
 A_\eta^2\zeta\!\left(\frac52-\eta\ii\rho,N\right)
 +2A_\eta B_\eta\zeta\!\left(\frac52+\eta\ii\rho,N\right)
 \notag\\[-1mm]
 &\hspace{42mm}
 +B_\eta^2\zeta\!\left(\frac52+3\eta\ii\rho,N\right)
 \bigg].
 \label{eq:J-leading-tail}
\end{align}
At fixed nonzero $\rho$ this improves the formal remainder to
$\order(N^{-5/2})$.  It is an asymptotic acceleration, not a rigorous
a posteriori error bound.  The separated endpoint coefficients are also
ill-conditioned as $\rho\to0$, where the two powers coalesce into logarithms.

\subsection{Physical combination before the outer sum}

For real $\lambda$ and $\rho$ the exact conjugation identities are
\begin{equation}
 I_{s,n}^{(-)}=\overline{I_{-s,n}^{(+)}},\qquad
 \overline{\omega_{-s}}=\e^{-s\pi\lambda}\omega_s.
 \label{eq:J-conjugation-identities}
\end{equation}
Let
\begin{gather}
 p=\frac52+\ii\rho,\qquad
 P=\frac{\Gamma(p)}{(-2\ii)^p\lambda^2},\qquad
 q_n=\frac{(p)_n}{n!},
 \label{eq:J-paired-prefactors}\\
 W_n=\sum_s\omega_sI_{s,n}^{(+)},\qquad
 V_n=\sum_s\e^{-s\pi\lambda}\omega_sI_{s,n}^{(+)}.
 \label{eq:J-WV}
\end{gather}
Define $P_-=\Gamma(\bar p)/[(-2\ii)^{\bar p}\lambda^2]$.  The ratio of the
conjugated minus prefactor to $P$ is
\begin{equation}
 R=\frac{\overline{P_-}}{P}=\e^{-\ii\pi p}=-\ii\e^{\pi\rho}.
 \label{eq:J-prefactor-ratio}
\end{equation}
Absolute convergence permits the physical subtraction to be performed for
each $n$ before the outer sum:
\begin{equation}
 J_{\lambda,\rho}=P\sum_{n=0}^\infty q_n
 \left(W_n^2-RV_n^2\right).
 \label{eq:J-paired-series}
\end{equation}
Choose $\chi=\e^{-\ii\pi p/2}$, so that $\chi^2=R$.  The summand is evaluated
without first forming two large squares as
\begin{equation}
 q_n(W_n-\chi V_n)(W_n+\chi V_n).
 \label{eq:J-factored-summand}
\end{equation}
This rearrangement avoids the final subtraction of independently accumulated branch sums
and thereby reduces the accumulated error, but it cannot remove the intrinsic conditioning
of the small factor $W_n-\chi V_n$.

The endpoint tail can be combined in the same fashion.  Here a plus subscript
denotes $\eta=+1$.  Write $A=A_+$, $B=B_+$ and
\begin{equation}
 A_V=\sum_s\e^{-s\pi\lambda}\omega_sL_{s,+},\qquad
 B_V=\Gamma(2+2\ii\rho)
 \sum_s\e^{-s\pi\lambda}\omega_sM_{s,+}.
 \label{eq:J-tail-AVBV}
\end{equation}
Then the physical leading tail is
\begin{align}
 \Delta_NJ\sim\frac{(-2\ii)^{-p}}{\lambda^2}\bigg\{
 &(A-\chi A_V)(A+\chi A_V)
 \zeta\!\left(\frac52-\ii\rho,N\right)
 \notag\\
 &+\left[(A-\chi A_V)(B+\chi B_V)
 +(A+\chi A_V)(B-\chi B_V)\right]
 \zeta\!\left(\frac52+\ii\rho,N\right)
 \notag\\
 &+(B-\chi B_V)(B+\chi B_V)
 \zeta\!\left(\frac52+3\ii\rho,N\right)\bigg\}.
 \label{eq:J-paired-tail}
\end{align}
The plotted $N=64$ result retains $n=0,\ldots,63$ in
eq.~\eqref{eq:J-paired-series} and adds eq.~\eqref{eq:J-paired-tail} beginning
at $n=64$.

\section{Fixed-mixing large-mass JWKB expansion}
\label{app:large-rho}

This appendix derives the fixed-$\lambda$ expansion quoted in
eqs.~\eqref{eq:C-large-rho}-\eqref{eq:theta-large-rho}.  The logarithms and
integration cycles are inherited from the Bunch-Davies prescription;
they are not re-principalized after a contour deformation.

\subsection{Soft coefficient}

For $\nu=\ii\rho$, the finite branch sum in eq.~\eqref{eq:Cplus} can be
reduced by gamma-function duplication and reflection identities to
\begin{equation}
 C_{\lambda,\rho}
 =\frac{\sinh(\pi\lambda/2)}{\sqrt{\pi}}\,
 2^{2\ii\rho}\Gamma(-2\ii\rho)
 \frac{
 \Gamma(\frac34+\frac{\ii\rho}{2})
 \Gamma(\frac34-\frac{\ii\rho}{2})}{
 \Gamma(\frac34-\frac{\ii}{2}(\rho-\lambda))
 \Gamma(\frac34-\frac{\ii}{2}(\rho+\lambda))}.
 \label{eq:C-exact-reduced}
\end{equation}
Stirling expansion of
the Gamma functions in eq.~\eqref{eq:C-exact-reduced} gives eq.~\eqref{eq:C-large-rho}.  Its leading
magnitude is proportional to $\e^{-\pi\rho}$.

\subsection{Reduced cycle, finite saddles, and branch cuts}

Let $S=1+t_1+t_2$.  The exact hard integral is the reduced double integral
\eqref{eq:J-double-integral}, with the two branches $J^\eta$ kept separate
until the final combination.  We fix $\ln (-2\ii)$ as in
eq.~\eqref{eq:J-log-branch} and continue $\ln  S$ from $S>0$.
These are separate logarithms: replacing them after deformation by a newly
principalized $\ln (-2\ii S)$ can move the integrand to a
different sheet.

The branch loci of the reduced representation are $t_j=0,-1,\infty$ and
$S=0$.  A useful branch-safe representative of the positive cycle is
\begin{equation}
 t_j(r)=r+\ii\varepsilon\frac{r}{1+r},\qquad
 r\geq0,\qquad \varepsilon>0,\qquad \varepsilon\to0^+.
 \label{eq:large-rho-lateral-cycle}
\end{equation}
For $r>0$, this contour reaches neither the negative-real cuts of $t_j$ and
$1+t_j$ nor the cut of $S$; moreover, $\operatorname{Im}S>0$.  At $r=0$ it
ends at the inherited endpoint branch point without crossing a cut.

To identify the JWKB sectors,
we first notice that the integral is symmetric for both $t_1$ and $t_2$, so we can solve the same differential equation for the contour without loss of generality. Set
\begin{equation}
 t=\sinh^2\frac{\chi}{2}.
 \label{eq:large-rho-conical-coordinate}
\end{equation}
The exact equation for each $h_s$ in eq.~\eqref{eq:h-numerical} becomes
\begin{equation}
 \frac{\dd^2h_s}{\dd\chi^2}
 +\coth\chi\frac{\dd h_s}{\dd\chi}
 +\left(\rho^2+\frac14
 +\lambda^2\operatorname{csch}^2\chi\right)h_s=0.
 \label{eq:h-conical-equation}
\end{equation}
Away from $\chi=0$, its two JWKB waves are proportional to
$(\sinh\chi_i)^{-1/2}\e^{\ii\sigma_i\rho\chi_i}$, with $\sigma_i=\pm1$.  For a
two-leg sector (denoted by subscripts $1$ and $2$), the large phase is
\begin{equation}
 \Phi_{\boldsymbol\sigma}^{\eta}
 =\sigma_1\chi_1+\sigma_2\chi_2
 -\eta\ln  S.
 \label{eq:large-rho-JWKB-phase}
\end{equation}
Varying the two variables independently gives
\begin{equation}
 \sqrt{t_j(1+t_j)}=\eta\sigma_jS,\qquad
 (t_1-t_2)S=0.
 \label{eq:large-rho-stationarity}
\end{equation}
The factor $S=0$ is the branch divisor, not a regular saddle.  The complete
regular finite pair is therefore
\begin{equation}
 t_1=t_2=-\frac12+\frac{\ii\tau}{2\sqrt3},\qquad
 \tau=\pm1.
 \label{eq:large-rho-finite-saddles}
\end{equation}
On the logarithmic lift continued from the positive cycle,
\begin{equation}
 \chi_j=\eta\sigma_j\left(\frac12\ln 3
 +\frac{\ii\tau\pi}{2}\right),\qquad
 \Phi_\tau^\eta=\eta\left(\frac32\ln 3
 +\frac{\ii\tau\pi}{2}\right).
 \label{eq:large-rho-saddle-action}
\end{equation}
Thus the member with $\eta\tau=-1$ would grow exponentially and has zero
intersection with the inherited positive cycle.  The member with
$\eta\tau=+1$ is exponentially damped and does not change the algebraic
endpoint expansion derived below.

The local descent directions are also fixed by the transported cycle.  Put
$\kappa=\sigma_1\sigma_2$ and
\begin{equation}
 v_{(1)}=\frac{(1,\kappa)}{\sqrt2},\qquad
 v_{(2)}=\frac{(1,-\kappa)}{\sqrt2}.
\end{equation}
At either finite saddle the Hessian in the $\chi_j$ coordinates has
eigenvalues $3\eta/2$ and $-\eta/2$ along these two vectors.  A local
steepest-descent patch for $\e^{\ii\rho\Phi}$ is
\begin{equation}
 \delta\boldsymbol\chi
 =\e^{\ii\eta\pi/4}r_{(1)} v_{(1)}
 +\e^{-\ii\eta\pi/4}r_{(2)} v_{(2)},\qquad
 r_{(1)},r_{(2)}\in\mathbb R.
 \label{eq:large-rho-descent-directions}
\end{equation}
The choices $r_{(j)}\to-r_{(j)}$ are the two arms of the same oriented descent line. 

In the principal-cut $t$-chart and with the flat $\chi$-metric used here, continuation of the pure Hessian downward arms reaches the boundary of that chart, where $t_j$, $1+t_j$, or $S$ lies on its negative-real cut. Hence these unbroken arms
cannot by themselves replace the inherited positive cycle on the same sheet.
A full infinite descent ray therefore cannot
replace the positive-real cycle without explicit cut-lip contributions.  One
may instead join a truncated local descent patch to the branch-safe cycle
\eqref{eq:large-rho-lateral-cycle}.  For the algebraic large-$\rho$ terms no
such deformation is required, because the leading region is the positive
endpoint itself.

\subsection{Endpoint Bessel matching}

Near $\chi=0$, eq.~\eqref{eq:h-conical-equation} is a conical Bessel problem.
Set $u=\rho\chi$ and $L=\pi\lambda/2$.  Combining the two branches before taking the endpoint limit gives
\begin{align}
 \mathcal H_{\lambda,\rho}\!\left(
 \sinh^2\frac{u}{2\rho}\right)
 &=F_\lambda(u)+\order_\lambda(\rho^{-2}),
 \label{eq:large-rho-endpoint-profile}\\
 F_\lambda(u)
 &=\frac{\ii}{2\cosh L}
 \left[\e^{-L}J_{-\ii\lambda}(u)
 -\e^LJ_{\ii\lambda}(u)\right]
 =-\ii\e^{-L}\sinh L\,H^{(1)}_{\ii\lambda}(u).
 \label{eq:large-rho-Flambda}
\end{align}
Here $J_\alpha$ and $H_\alpha^{(1)}$ denote the Bessel and Hankel functions.
The endpoint scale is therefore $\chi=\order(\rho^{-1})$, or
$t=\order(\rho^{-2})$.  The required integrals are
\begin{equation}
 \int_0^\infty\dd u\,uF_\lambda(u)=\lambda,\qquad
 \int_0^\infty\dd u\,u^3F_\lambda(u)
 =-\lambda(\lambda^2+4).
 \label{eq:large-rho-Abel-moments}
\end{equation}
These are generalized oscillatory moments defined by Abel, or equivalently
Mellin, continuation and are not absolutely convergent ordinary integrals.
For example, they follow by analytic continuation of
\begin{equation}
 \int_0^\infty\dd u\,u^\mu J_\alpha(u)
 =2^\mu\frac{
 \Gamma\bigl(\frac{\alpha+\mu+1}{2}\bigr)}{
 \Gamma\bigl(\frac{\alpha-\mu+1}{2}\bigr)}.
 \label{eq:large-rho-Bessel-Mellin}
\end{equation}

For the minus branch, denote the double integral without its gamma prefactor
by
\begin{equation}
 \mathcal I_-=\int_0^\infty\dd t_1\int_0^\infty\dd t_2\,
 \mathcal H_{\lambda,\rho}(t_1)
 \mathcal H_{\lambda,\rho}(t_2)S^{-p_-}.
 \label{eq:Jminus-endpoint-integral-definition}
\end{equation}
In the endpoint layer,
\begin{align}
 \dd t&=\frac{u\,\dd u}{2\rho^2}+\order(\rho^{-4}),\\
 S^{-p_-}
 &=1+\frac{\ii}{4\rho}(u_1^2+u_2^2)
 +\order(\rho^{-2}).
\end{align}
Using eq.~\eqref{eq:large-rho-Abel-moments} gives
\begin{equation}
 \mathcal I_-=\frac{\lambda^2}{4\rho^4}
 \left[1-\frac{\ii(\lambda^2+4)}{2\rho}
 +\order_\lambda(\rho^{-2})\right].
 \label{eq:Jminus-endpoint-integral}
\end{equation}
On the branch \eqref{eq:J-log-branch}, Stirling expansion of the remaining
prefactor gives
\begin{equation}
 \frac{\Gamma(p_-)}{(-2\ii)^{p_-}}
 =\frac{\sqrt\pi}{4}\rho^2
 \exp\!\left[-\ii\rho\left(\ln \frac{\rho}{2}-1\right)
 +\frac{\ii\pi}{4}\right]
 \left[1+\frac{47\ii}{24\rho}+\order(\rho^{-2})\right].
 \label{eq:Jminus-gamma-prefactor}
\end{equation}
Multiplication of the last two equations yields
\begin{equation}
 \begin{aligned}
 J^-_{\lambda,\rho}
 ={}&\frac{\sqrt\pi\,\lambda^2}{16\rho^2}
 \exp\!\left[-\ii\rho\left(\ln \frac{\rho}{2}-1\right)
 +\frac{\ii\pi}{4}\right]
 \\[-1mm]
 &\times\left[1-\frac{\ii(12\lambda^2+1)}{24\rho}
 +\order_\lambda(\rho^{-2})\right].
 \end{aligned}
 \label{eq:Jminus-large-rho}
\end{equation}
The $J^+$ endpoint carries one additional factor $\e^{-\pi\rho}$.  Therefore
$J=J^+-(J^-)^*$ gives eq.~\eqref{eq:J-large-rho}.  Combining it with
eq.~\eqref{eq:C-large-rho} gives eqs.~\eqref{eq:CJ-large-rho} and
\eqref{eq:theta-large-rho}.  Terms from the damped finite saddle, if present,
are beyond all algebraic orders displayed here.

\bibliographystyle{JHEP}
\bibliography{biblio}

\end{document}